\documentclass[12pt]{article}
\usepackage{amsmath,amssymb,latexsym}
\usepackage{graphicx}

\begin{document}
\Large

\begin{center}
Superbubbles,  Galactic Dynamos  and  the Spike Instability
\vskip .2 in
Russell M. Kulsrud
\vskip .2 in 
Princeton Plasma Physics Laboratory 
\end{center}
\vskip .2 in
\normalsize

\section{ Abstract}
We  draw attention to a problem with the alpha-Omega 
dynamo when it is applied to the origin  of the galactic 
magnetic field under the assumption of perfect flux freezing.
The standard theory involves the expulsion of undesirable
flux and, because of flux freezing,  the mass anchored on this flux also 
must be expelled.  The strong galactic gravitational field 
makes this  impossible on energetic grounds.
  It is shown that if only   short pieces
of the undesirable field lines are  expelled, then mass 
can flow  down along these  field lines without requiring much energy.
This expulsion of only short lines of force can be accomplished
by a spike instability associated with gigantic astrophysical superbubbles.
The physics of this instability is discussed and the  results enable 
an estimate to be made  of the number of spikes in the galaxy. It appears
 that there are probably enough spikes to cut all the undesirable lines
into pieces as short as a couple of  kiloparsecs  during a dynamo time
of a billion years. These cut pieces then may be randomly rotated
in a dynamo time  by alpha-Omega  diffusion and there
is enough rotation  to  get rid of the undesirable  flux without
expelling the fields themselves.  The spike process seems
strong enough to allows the 
alpha-Omega dynamo to create the galactic field without
any trouble from the boundary condition problem.

\section{ Introduction}
Our galaxy is believed  to have  finite magnetic flux
in the form of a toroidal field, because this field does
not reverse across the galactic midplane.
The origin  of this field is paradoxical.
This is because, on the galactic scale, flux freezing
is almost infinitely strong so that  flux through
 any moving plasma region cannot change.  In fact,
the change is only finite over a Hubble time if the
scale of variation is smaller than an astronomical unit.
Thus, the obvious question arises:  How can one start
with a weak field of cosmological origin  and increase it 
to its present value?

The answer is: that the flux in the galactic disc
itself need not be constant,  but only the flux  in the larger
region consisting of the disc plus the galactic halo.
The standard alpha-Omega theory (Steenbeck et. al [1])
supposes one starts with a very 
weak  field  whose origin   is 
cosmological,  and is  amplified by compression 
during galactic formation.  (Such an initial  field might
have a strength of $ B_0 =10^{ -12 } $ gauss,  be
 toroidal in the positive direction,
and  uniformly fill the disc with flux $ \Phi  $.)
The standard alpha-Omega dynamo 
folds this field  back and forth by the alpha effect.
  The result is  a flux of 
  $ 2 \Phi $   in the positive direction
 about the midplane  and a negative
 flux of $ - \Phi $ near the edges of the disc.
Then the negative flux is supposed to be 
turbulently diffused  out of the
disc and into the halo leaving the disc with double  its
original flux.  This doubling  takes place during a
'dynamo time'  of  less than a billion years.
It repeats over and over perhaps  twenty
times in the life of the galactic disc amplifying 
the field strength (and flux  in the disc) by over a million
times. Such a process appears to provide a  reasonable
origin  of our present galactic field, and also satisfies
the magnetic flux freezing condition,  (Parker [2],
Ruzmaikin, et. al. [3]).

However, when one looks more closely at the diffusion 
 of flux from the disc to the halo, a problem
arises because the galaxy has a very strong
gravitational  field.  In  the motion of a flux
tube into the halo,  the mass anchored onto it  by flux freezing
 must also
be lifted into the halo.

During this motion an energy,  equivalent  to an escape  velocity of
four hundred kilometers a second, must be supplied to the mass.
  This energy
is extremely large compared to  energies in 
interstellar turbulence, which have typical velocities 
of ten kilometers a second.  Where can such energies
come from?  

One suggestion is supernovae.   But velocities of supernovae remnants 
 are reduced to a few tens of kilometers a second
by   snow plowing  all the surrounding interstellar mass.
   They have only a small chance of breaking out of the
galactic disc.

An alternative suggestion is the   recently  recognized phenomena 
of superbubbles, which are driven by multiple supernova.
Some of these actually break out of the galactic disc.
 However, by the time they leave the disc,
the mass that they have snow plowed has slowed  them down
to a velocity of fifteen to twenty kilometers a second.
 So this can not be a direct source of the large
 energies needed for diffusion
of the tubes into the halo.

One is forced to consider a different concept. This concept 
is based on the realization  that the flux freezing process
applies only to motions perpendicular to the line 
of force.  It in no way constrains the parallel motion
along the field.  Thus,  if a 
 short length $ \ell_S $  of a line of force 
is  lifted
out of the disc and into the halo,    the mass anchored on it
 can be greatly
reduced from  its initial value
by  sliding  down along 
a length $ \ell_V $  to the rest of the line $ \ell_R $ 
remaining in the disc. 
Because of the small mass on the $ \ell_S $  little energy
is required.  Thus,  on energy grounds alone,
such a process is possible.  But removing a short piece of a 
negative line
into the halo does not itself  provide  a  reduction in the negative 
 toroidal flux in the disc. A further idea is needed (Kulsrud [4]).

Consider what happens to the rest of the negative field line $ \ell_R $
 on either
side of $ \ell_S  $?  There is  a gap in it caused by the
removal of $ \ell_S $.  Of course, 
 this gap is kept closed by two vertical 
pieces of the field $ \ell_V $,  
connected to $ \ell_s $.  But the field strength of
these  pieces, $ \ell_V $, becomes very weak by horizontal 
expansion. The region into
which these $ \ell_Vs $ expand is much larger than the
thickness of the disc by  a factor of  fifty,  the aspect ratio of the disc,
  so  that the field strength of these $ \ell_V $ become 
 far too weak to affect  the rest of the line, $ \ell_R $.
 This part of the line  acts as though  its two pieces are effectively
decoupled or cut. Therefore,  the ends of $ \ell_R $ are free
to move and it appears that this part of the field line
has free ends.  

Suppose a given line of force is 'cut' into a number
of pieces by  removal of a number of 
 short   $ \ell_S $  pieces. 
Then, when these  $ \ell_{Rs} $ are  acted on by $ \beta $, 
 the turbulent diffusion
of the alpha-Omega dynamo,  they
will be rotated  randomly in direction and no longer preserve their
negative toroidal  flux. The rotations
transfer their negative flux to the $ \ell_S $  pieces in the halo.
Due to the weakness of the $ \ell_V $ pieces
this transfer through the weak $ \ell_V $ pieces of flux
will have no dynamic  effect on the disc pieces $ \ell_R $.
But this transfer effectively removes negative
flux  into the halo as required by the alpha-Omega  dynamo.
Only  a small  amount of energy is required for these processes.
See figure 1.

\begin{figure}
\rotatebox{0}{ \scalebox{0.55}{ \includegraphics{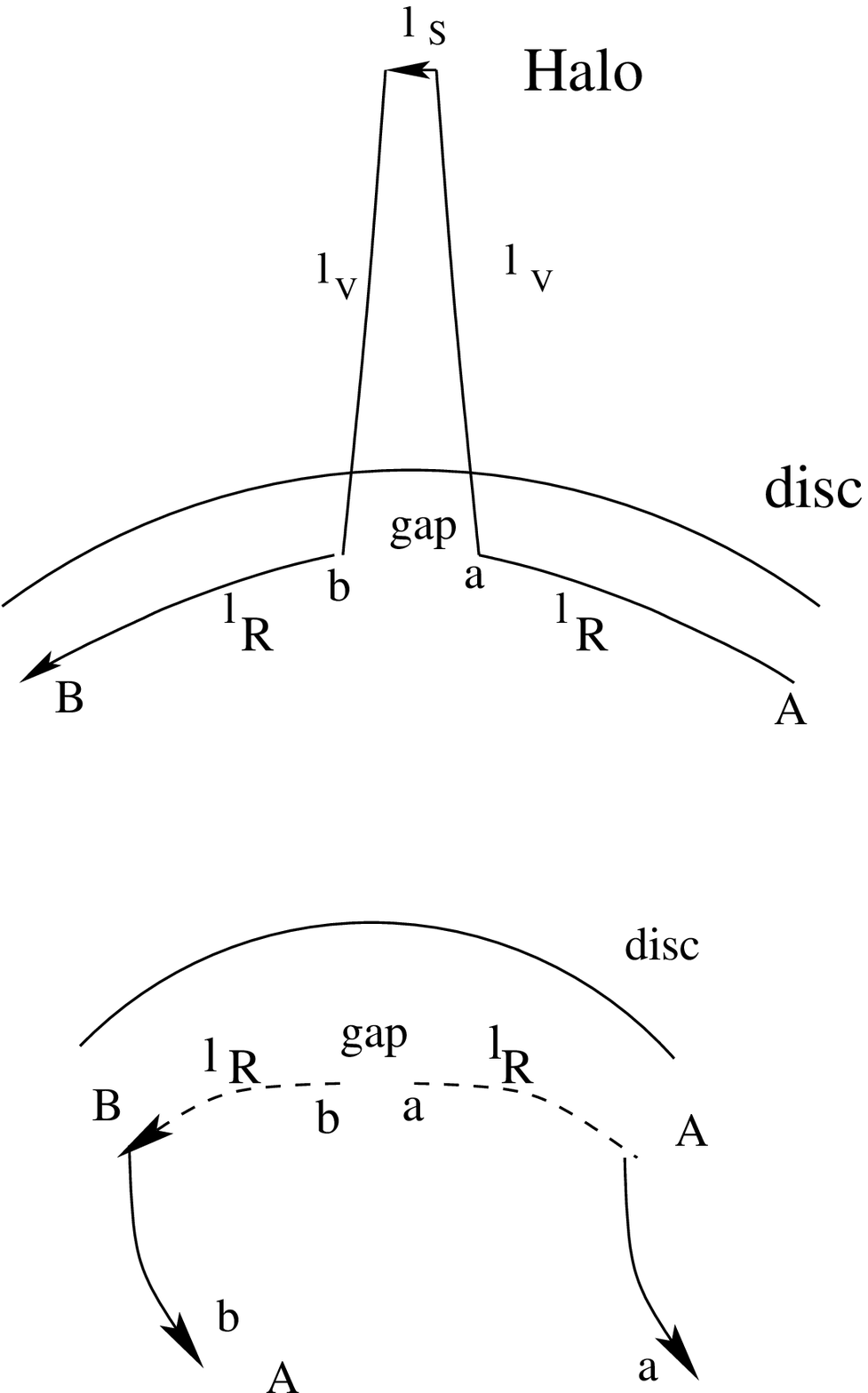} } }
\caption{(a) A short  piece of a field line $ \ell_S $ is lifted into the
halo.  It is connected to the pieces $ \ell_R $ in the disc
by weak pieces $ \ell_V $. (b) There is a gap in $ \ell_R $ that allows the 
ends of the field lines to rotate freely under diffusion  so the points
$ a $ and $ b $ rotate to new positions.}
\end{figure}
\vspace*{1in}

It remains to discover a mechanism  to take the small 
$ \ell_S $ pieces  and  propel them far into the halo.
In paper I (Kulsrud [5]) it is     proposed that superbubbles
 present a way to do this.

Each  superbubble arises
from the formation of a normal galactic star cluster.
In the star cluster the stars are formed in  a short
time and the more massive O and B stars in the cluster
continually evolve  and
explode into supernovae.  The many supernova explosions
overlap and their energy is transformed into a high
pressure low density plasma. Its pressure 
 forces all the surrounding interstellar 
matter into a gigantic rapidly expanding shell which
comprises the superbubble phenomena.  The shell gradually
slows down as all the surrounding mass accumulates
onto  it.  The shells of the  more 
luminous superbubbles manage to escape out of the disc.
Once they leave the disc, their expansion  accelerates
because the pressure of the core remains large
(the supernova are still exploding), and  there
is no more mass to snowplow. 

This 
situation where low density plasma is accelerating the high
density shell 
is unstable to the Raleigh-Taylor instability, (MacLow and McCray [6])
and the shell breaks up into many fragments.  
The fragments may still be connected and the pressure still
contained by the fragmented shells. This  leads to a secondary
instability the 'spike instability'. 

This instability is described in paper I. 
The spike is a narrow elongated eruption on a fragment.
The matter at the top of the spike is accelerated
against gravity,  and because of this gravity the mass
falls down  along the sides of the spike.
(See figure 2). As the mass at the top slides down
the top becomes lighter and, as long as the pressure
is confined,  it  accelerates faster.  This increases
the steepness of the sides of the spike and 
the down flow increases, making the top still lighter and
 accelerate faster.  This process I call the 'spike 
instability'. As the spike rises towards the halo it 
increases its speed until finally the velocity 
exceeds the escape velocity and the top mass  reaches
the halo ballistically.

\begin{figure}
\rotatebox{0}{ \scalebox{0.55}{ \includegraphics{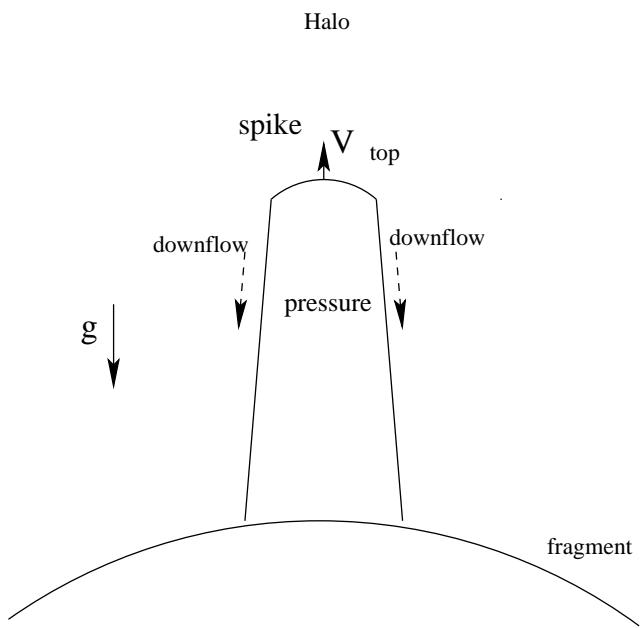} } }
\caption{As the spike grows, gravity $ g $ leads to
a downflow of the mass from the top of the spike to the fragment,
making the top lighter.  The superbubble pressure then lifts the top
faster}
\label{Fig1}
\end{figure}
\vspace*{1in}

The shell mass contains the interstellar field and
as the spike rises from the fragment, it carries the $ \ell_S $ pieces 
of the same lines of force 
with it.  Most of the lines slide down with the mass
but some of those  at the very top of the spike
 reach the halo
as desired for the model of the dynamo.

For each spike the $ \ell_R $ parts of these
lines of force  are cut.
When  enough lines are cut into short enough
pieces  the negative flux
can be diffused away by random turbulent rotation of these
$ \ell_R $ pieces.   If this occurs during a short enough time
compared to dynamo times (say a  billion years),  then the 
alpha-Omega dynamo should work and the galactic field 
amplified from a weak field to its present value.

\section{ The Spike Instability}

In this section we develop the theory of the spike instability
in order to determine  what fraction of the magnetic lines of force initially
in the spike are  expelled to the halo.

Then  we estimate the number of spikes 
and the average length of the cut lines, 
and finally,  we  calculate the amount  of angular randomization
 of the $ \ell_R $ lines that occurs during a dynamo time. 
It from this calculation it
appears that  the process of flux
removal works during  a dynamo time of a billion years.
In the final section, we summarize and draw the conclusions.

As a model for the spike   we approximate  the surface of the superbubble
fragments as  infinitely thin plane  sheets 
with surface mass $ \sigma_0 $ 
supported against gravity by a constant underlying pressure.
  Its surface density and actual
thickness $ D $ are derived in the paper I.  The gravitational 
acceleration $ g $  is the sum of the  true  gravity 
due to the stars in the disc and  the effective gravity from
the acceleration of the superbubble shell.  However,
in our  examples  the astrophysical gravity is the more important.

We work in cylindrical coordinates with $  \hat{{\bf z }}  $
vertical to the sheet 
and $  \hat{{\bf r }} \mbox{ and }  \hat{{\bf \theta  }} $ 
in the initial plane of the sheet. We write the
equations of the perturbed motion in a Lagrangian form. 
We assume cylindrical symmetry and assume  all quantities
depend only on  time $ t $ and  the initial radius
of the perturbation $ r_0 $. Let $ \xi_r(t,r_0) \mbox{ and } 
 \xi_z(t,r_0) $ be the components of the displacement.

Let the coordinates of the four corners of an undisplaced
 small square $ {\bf  A_0}  $ at time $ t = 0 $ be 
\begin{eqnarray} 
P1 &=&  [r_0, 0, 0] \\ \nonumber 
P2 &=&  [r_0+\delta ,0, 0], \\ \nonumber  
P3 &=&  [r_0,\epsilon, 0] \\ \nonumber 
P4 &=&  [r_0+ \delta ,\epsilon ,0]  
\end{eqnarray} 

  Then after the displacement,
the coordinates of the  four corners of 
the displaced square $ {\bf  A } $ are
\begin{eqnarray} 
P1 &=& [r_0 + \xi_r(r_0), 0, \xi_z(r_0)]:\\ \nonumber 
P2 &=& [ r_0 +\delta + \xi_r(r_0) +
\delta  \frac{\partial \xi_r(r_0)}{\partial r_0}, 0,
\xi_z(r_0) + \delta \frac{\partial \xi_z(r_0)}{\partial r_0}]  , \\ \nonumber 
P3 &=& [r_0 + \xi_r(r_0), \epsilon , \xi_z(r_0)] \\ \nonumber 
P4 &=& [ r_0 +\delta + \xi_r(r_0) +
\delta  \frac{\partial \xi_r(r_0)}{\partial r_0}, \epsilon ,
\xi_z(r_0) + \delta \frac{\partial \xi_z(r_0)}{\partial r_0}]  , 
\end{eqnarray}

 For notational convenience, let us 
 abbreviate the components of  the displacement as
\begin{eqnarray} 
\xi_r(r_0) &=& R \\ \nonumber
\xi_z(r_0) &=& Z
\end{eqnarray}
After the displacement, the components of the
 vector $ {\bf  A } $  are
\begin{eqnarray} 
A_r &=&  -\epsilon \delta r Z' = -\epsilon \delta (r_0+R) Z' \\ \nonumber 
A_z &=&  \epsilon \delta r (1+ R')=
  \epsilon \delta (r_0+R)(1+R') 
\end{eqnarray}
where prime denotes a derivative with respect to $ r_0 $.
(Where convenient we drop the arguments $ t, $ and $  r_0 $).
The vector $ \bf{A} $ is perpendicular to the displaced square
and its magnitude is equal to its area.

In the equation of motion of the surface of the spike
the mass of the square is $ \sigma_0 |{\bf A_0}|$,
the pressure force on it is  $ p \bf{A} $,  and the
gravitational force is $ -g \sigma_0 |{\bf A_0}|  \hat{{\bf z }} $. 
From the initial equilibrium we have $ g \sigma_0 = p $ 

Thus, the equations of motion of the displacement are 
\begin{eqnarray} \label{eq:A}  
\frac{\partial^2 R}{\partial t^2} &=& 
- g ( 1+ \frac{R}{r_0} ) Z' \\ \nonumber 
\frac{\partial^2 Z }{\partial t^2} &=& 
 g ( 1+ \frac{R}{r_0} ) (1+R') -g 
\end{eqnarray} 
where the prime denotes the derivative with respect to 
$ r_0 $.

For the linear solution of these equations
assume that $ R $ and $ Z $ are proportional to 
$ \exp{ \gamma t} $.  Keeping the linear terms we have
\begin{eqnarray} 
\gamma^2 R &=& - g Z'  \\ \nonumber 
\gamma^2 Z &=& = g (\frac{R}{r_0}+ R')
\end{eqnarray} 

Defining 
\begin{equation}
k = \frac{\gamma^2}{k g}
\end{equation}
we find 
\begin{eqnarray}
Z''+ \frac{Z'}{r_0} + k^2 Z &=& 0 \\ \nonumber 
R''+ \frac{R'}{r_0} + k^2 R - \frac{R}{r_0^2 }=0
\end{eqnarray}
so the linear solutions  regular at $ r_0 = 0 $ are
\begin{eqnarray} 
Z &=&  a e^{\gamma t} J_0(k r_0) \\ \nonumber
R &=& a e^{\gamma t}J_1(k r_0) 
\end{eqnarray}
The growth rate is $ \gamma= \sqrt{ k g} $.
The shortest wave length consistent with an equilibrium  sheet mass
is given by $ k D = 1 $ where $ D $ is the actual thickness
of the layer which we approximated above as a sheet. 
An estimate of this thickness, which is given in 
paper I, is about a  parsec.  The value of $ g $  in the neighborhood 
of the disc is  $ 10^{ - 8 } \mbox{ cm} /\mbox{ sec}^2 $.
These values give  the linear time of growth,  $ t_G = 1/\gamma  =
  7.4 \times 10^{ 5} $ years, a time short compared to the
life time of the superbubble (usually 20-50 million years).

To get the long time behavior of the spike, we approximate
Equation~(\ref{eq:A})   by its nonlinear terms.
\begin{eqnarray} \label{eq:B}
\frac{\partial^2 R }{\partial t^2} &=& 
- \frac{g}{r_0 } R Z' \\ \nonumber 
\frac{\partial^2 Z }{\partial t^2} &=& 
\frac{g}{r_0} R R' 
\end{eqnarray} 

These equations have a solution with $ R $ and $ Z $ proportional
to $ 1/(t_0 - t)^2 $. Setting 
\begin{eqnarray} 
R &=& \frac{\eta }{(t_0-t)^2} \\ \nonumber   
Z &=& \frac{\nu }{(t_0-t)^2} 
\end{eqnarray} 
We have from the first equation, 
\begin{equation}
  6 \eta  = - \frac{g}{r_0} \eta \nu'
\end{equation}

Canceling $ \eta  $ from this equation,  and solving for $ \nu $ gives
\begin{equation} 
\nu= C-3 \frac{r_0^2}{g}
\end{equation}
The  equation for $ \eta $ is
\begin{equation}
6 \nu = \frac{g}{r_0} \eta \eta'
\end{equation}
which on substituting   the solution for $ \nu $  gives
\begin{equation} 
(\eta^2)'= \frac{12 C}{g} r_0 - \frac{36}{g^2} r_0^3
\end{equation}
and 
\begin{equation}
\eta^2 = \frac{6 C}{g} r_0^2 - \frac{9}{g^2}r_0^4
\approx \frac{6 C}{g} r_0^2
\end{equation}

The constants $ C $ and $ t_0 $ are determined by
matching both  terms of 
the solution  for $ Z $ to its  linear solution at $ t= 0 $. 
\begin{equation}
\frac{(C- 3 r_0^2/g)}{t_0^2} = a(1-k^2r_0^2/ 4) 
\end{equation}
which gives
\begin{eqnarray} 
\frac{3}{g t_0^2}  &=&  \frac{k^2 a}{4 } \\ \nonumber 
\frac{C}{t_0^2} &=& a
\end{eqnarray}

Solving  these two equations for $ t_0 \mbox{ and } C $
we have
\begin{eqnarray} 
t_0  &=& \sqrt{ \frac{12}{k^2 g a }} \\ \nonumber 
C &=&  a t_0^2 = \frac{12}{k^2 g} 
\end{eqnarray} 
so 
\begin{equation} \label{eq:C} 
Z= \frac{12}{k^2 g} \frac{(1-k^2 r_0^2/4)}{(t_0-t)^2} 
\end{equation}

Treating $ R $ the same way we get
\begin{equation}
R == \frac{\sqrt{ 72}}{k g} \frac{r_0}{(t_0-t)^2} 
\end{equation}

Now let us determine the time, $ t_e $ ,   at which the upward
velocity of the peak of the spike reaches the escape velocity
from the of the disc, $ v_e $, (Heiles [7]).  That is,
from  $ \partial Z/\partial t =v_e $, (Equation~(\ref{eq:C})),  
\begin{equation} 
\frac{\partial Z}{\partial t } = \frac{24  }{k^2 g (\Delta  t)_e^3} 
= v_e
\end{equation}
we get
\begin{equation} 
(\Delta t)_e = t-t_e = \left( \frac{24}{k^2 g v_e}\right)^{1/3}
\end{equation}  .

This occurs at the height 
\begin{equation} 
h_e= 1.44 \left( \frac{v_e^2}{k^2 g} \right)^{1/3}
\end{equation}
With the above numbers  $ h_e = 49 $ parsecs.

\section{ The Number of Lines Cut by One Spike}

To determine the number of lines expelled  we start
with the number of lines initially embedded in the spike.
This is equal to  the number of lines embedded in the 
superbubble shell in  
a strip with the width of one spike, $ \approx \lambda = 1/ k $
Since the shell consists of all the interstellar mass swept up by
the superbubble, the strip contains   $ B \lambda H $ 
lines,     where  $ B $ is mean  field in the disc.
 At the escape time $ t_e $, the region
 $ r_0  < \lambda $ has horizontally  expanded  a distance
\begin{equation}
\frac{R(t_e, \lambda ) }{r_0}=
\frac{\sqrt{ 72}}{kg}\frac{1}{(\Delta t)^2_e}
=\frac{72^{1/2}}{(24)^{2/3}} \left(\frac{k v_e^2}{g}\right)^{1/3} =
37.59 \mbox{pc} .
\end{equation}
 This expansion reduces the field
strength (i.e. the density of the lines) at the top
of the spike, by 
a factor of  
\begin{equation}
\frac{r_0}{R(t_e, \lambda ) }=.0266 
\end{equation} 

However, by the time the spike has reached a height of
$ h_e $, the solution of the spike has spread out a distance
of $ 37 $ parsecs, and the behavior of the spike
a distance $ r \gg  \lambda $ from the cylindrical axis is
uncertain.  Therefore, we  take for 
    the number of expelled lines, $ \Phi_e $, the number of lines
at the top of one spike, within  $ r_0< \tau \lambda $.  
(Those that are certainly expelled.)
\begin{equation}
\Phi_e = 2 \times .0266 \times BH \tau  \lambda 
\end{equation} 
The other lines may fall back down, or possibly be expelled.

Among the these lines there are both positive and negative
lines.  The proportions depend on  how high above the
galactic midplane the superbubble started.  
Also, the sheet of plasma, assumed in our derivation 
to be infinitely thin,  actually
has a finite thickness.  In the radial outflow flow, the mass in 
the sheet nearest the superbubble pressure will probably
be squeezed out faster than that farther away.  
Without analyzing this further we simply assume that  
half the number of lines at the top of the spike are negative
and ignore the positive lines.
Thus, we take  the number of negative  lines cut per spike
to be
\begin{equation}
\frac{1}{2} \Phi_e =  .0266 B H \lambda \tau 
\end{equation}
 
Denote   the number of spikes per fragment  as
$ f_S $,  and the number of fragments per superbubble 
as $ N_f $.  The size of the spike is $ \lambda \sim D $
where $ D $ is the thickness of the superbubble shell.

Consider a superbubble of luminosity $  \bar{L} =
1.37 \times 10^{ 37} $ ergs per second, This is  the luminosity
of a superbubble just strong enough to break out of the disc,  and
fragment. For such a superbubble we have, from paper I, 
that 
\begin{eqnarray} \label{eq:D} 
N_f &=&  \frac{78}{T^2_{300} } \\ \nonumber  
D &=& \frac{H}{222} T_{300} 
\end{eqnarray}
where the temperature, $ T=300 \times T_{300} $ degrees Kelvin, 
is the temperature in the superbubble shell.
$ T $  is generally assumed to be between a hundred and 
a thousand degrees  Kelvin.
Then our estimate  for the number of cuts of negative lines   
by a single  superbubble $ \Phi_{sb } $ is 
\begin{eqnarray} 
\Phi_{sb} &=& \frac{1}{2}  \Phi_e N_f f_S 
=.0266 B D H N_f f_S \tau 
\end{eqnarray} 

Substituting the numbers from Equation~(\ref{eq:D}) 
\begin{equation}
\Phi_{sb}= .0266  \frac{78}{222}  \frac{B H^2 f_S \tau  }{T_{300}}
=0.00935  \frac{ B H^2 f_S \tau }{T_{300}}
\end{equation}

Now what is the number of lines cut by a superbubble 
with luminosity $ L> \bar{L} $?
Again from paper I we find that $ N_f \sim (\bar{L}/L)^{2/3} $ 
and $ D \sim   (L /\bar{L})^{1/3} $ so the number of
negative lines cut by a  luminosity-$ L $ superbubble is
\begin{equation}
\Phi_L =  \Phi_{sb} \left( \frac{L}{\bar{L}} \right)^{-1/3} 
\end{equation}

Consider an annular region in the galaxy with mean radius 
$ R_S = 8.5 \mbox{ kpc} $ (the solar galactic radius)
and radial thickness $ \Delta R $. The rate of birth of superbubbles
with luminosity in $ d L $ is (Ferri\`{e}re [8]). 
\begin{equation}
d \sigma = 0.86 \times 10^{ -7} \left( \frac{L }{\bar{L} } \right)^{-2.3}
\frac{d L}{\bar{L} } \mbox{ kpc}^{-2} \mbox{ yr}^{-1} 
\end{equation}
and the  rate of birth in the entire annulus is
\begin{equation}
2 \pi R_S \Delta R d \sigma 
\end{equation} 
so that the rate of line cutting by all the superbubbles with
luminosity $ L $ 
in the annulus  is
\begin{equation}
\Phi_L d \sigma 2 \pi R_s \Delta R
\end{equation}

Now, the number of magnetic lines of force
 in the annulus, assuming the lines are
toroidal,  is $ 2 B H \Delta R $.  But only a quarter of these
lines are negative, so their number,  $ \psi_B $, is
\begin{equation} 
\psi_B = .5  B H \Delta R 
\end{equation}

The rate of cutting of any given line $ \rho_C $ by all 
the superbubbles with 
luminosity greater than $ \bar{L} $ is the rate
of cutting by all these superbubbles (integrated over their
luminosity) divided by the number
of lines, $ \psi_B $.
\begin{eqnarray}
\rho_C &=& \frac{2 \pi  R_s \Delta R \int_{L>\bar{L} } \Phi_L d \sigma \Phi_L}
{ .5 B H \Delta R} \\ \nonumber 
&=& 4 \pi  \times .00985 \times (.86 \times 10^{ -7}) R_S H
 \int^\infty_{L>\bar{L}} \frac{d L}{\bar{L} } 
\left( \frac{L}{\bar{L}} \right)^{-2.3-1/3}  \\ \nonumber 
&=& 4  \pi \times .00935 \times (.86 \times 10^{ -7} ) \times R_S H 
\frac{f_S \tau  }{T_{300} (1.3+1/3)} \\ \nonumber 
 &=& .17 \times 10^{ -7} \frac{f_S \tau }{T_{300}(1.3 +1/3)}
 \mbox{ yr}^{-1}   \\ \nonumber 
&=& 10.5 \frac{  f_S \tau }{T_{300}} \mbox{ cuts per billions years}
\end{eqnarray} 

Thus, in a billion years, each line of length $ 2 \pi R_S 
= 53 $ kiloparsecs will be cut  into $ 10.5  f_S \tau /T_{300} $ 
pieces of lengths $ \ell =5.1 T_{300}/f_S \tau  $ kiloparsecs.

These lines will be rotated through an angle $ \Delta \theta $ 
by the same turbulent diffusion $ \beta $ as included in the
normal alpha-Omega dynamo.  The rate of spread in angle  is
\begin{equation}
\frac{(\Delta \theta)^2 }{2 } = \frac{\beta t}{\ell^2}
\end{equation}
Take   $ \beta =\beta_{26} \times 10^{26} 
\mbox{ cm}^2/\mbox{ sec} $.   $ \beta_{26} \approx 0.5 $ 
(Ferri\`{e}re [8]).   
Then
\begin{eqnarray}
\Delta \theta &=&  0.52 \frac{\sqrt{ \beta_{26} t_9}}{\ell_{kpc}} \\ \nonumber 
 &=&  0.11  \sqrt{ t_9 } \frac{f_S \tau }{T_{300} } 
\end{eqnarray}
where $ t_9 $ is the time in billions of years and
$ \ell_{kpc} $ is the length of the cut pieces of lines
in kiloparsecs.

This, estimate  indicates that
the boundary condition on the alpha-Omega dynamo 
may  still be a problem.  The uncertainties of the estimate
are encapsulated in the parameters $ f_S, \tau , \mbox{and }
T_{300}  $.  If the temperature $ T $ in the
superbubble shell were   as  low as 100 degrees, and the parameter
$ \tau $  was as large as 3, then the random rotation $ \theta
$ would be of 0.9. This value for $ \theta $ is large enough to satisfy the
alpha-Omega galactic dynamo with a dynamo time of a billion years. This choice
of the parameters is quite reasonable.

It should be noted that, in this model for flux expulsion,
the galactic  magnetic field  has been assumed weak
enough that any magnetic tension is too weak to
 interfere with the spike instability.  This limit
on the field is reasonable since the main problem in the
origin  of our galactic magnetic field  is how to amplify
it when it is extremely weak.  When the galactic field
reaches its present value, presumably this back reaction
of the field will saturate the flux expulsion process 
and end any further growth of the field.

\section{ Conclusion}
The  purpose of this paper is to form a more precise estimate
of the flux expulsion by spikes associated  with superbubbles,
than was given in paper I. The
nature and physics of the spikes certainly 
introduces considerable   uncertainty into the conclusions
as to whether superbubbles can produce
the required destruction of negative flux needed for the dynamo 
to act on a time scale of a billion years.   In this note
we present  a conservative estimate of the superbubble 
process that shows that one {\it  should}.  
 take the spikes seriously in dynamo theory. A multi-dimensional
simulation of the spikes is clearly necessary to properly evaluate 
whether this process is  a reasonable way to make the alpha-Omega
dynamo viable for our galaxy.  However,
it should be noted that there are serious
difficulties with the dynamo itself, and at the moment there
does not seem to be any other reasonable alternatives.

\section{ Acknowledgments.}  I would like to acknowledge 
many useful  conversations with  my colleagues
about the feasibility of superbubbles in completing
the galactic dynamo theory.  These conversations have
given me enough confidence to publish these ideas.
This work
was carried out under the support of the Center of Magnetic 
Self Organization (CMSO) and under the US DOE contract No.
DE-AC-09CHI1466.

\section{References}
\begin{enumerate}
\item  Steenbeck, M., Krause, F., R\"{a}dler, K-H, 1966
Z. Naturforsch. {\bf  26a}, 369

\item  Parker, E N 1971
ApJ,  {\bf  138}, 552

\item  Ruzmaikin, A. A., Shukurov, A. M., and  Sokoloff, D. D.,
Magnetic  Fields in Galaxies, Kluwer Academic Publishers,
Dordrecht, Boston, London, 1988

\item Kulsrud, R. M. 2010,
Astronomich Nachrichten, {\bf  331}, 22

\item 
 Kulsrud, R. M. 2015,
submitted to J. Plasma Phys. (arXiv:1502.01712)

\item 
 MacLow,M.M., McCray, R., 1989, ApJ,{\bf  337}, 141

\item 
 Heiles, C. 1987 ApJ, {\bf  315}, 555

\item 
 Ferri\`{e}re, K. 1993, ApJ, {\bf  409}, 248

\end{enumerate}

\end{document}